\documentclass[conference]{IEEEtran}
\IEEEoverridecommandlockouts
\IEEEpubid{%
  \makebox[\textwidth][l]{%
    \parbox[t]{\columnwidth}{%
      \footnotesize\raggedright\setlength{\parskip}{0pt}%
      \textcopyright{} 2026 IEEE. Personal use of this material is permitted.
      Permission from IEEE must be obtained for all other uses, in any current or future media, including reprinting/republishing this material for advertising or promotional purposes, creating new collective works, for resale or redistribution to servers or lists, or reuse of any copyrighted component of this work in other works. \textbf{Presented in RAST 2026.}%
    }%
  }%
}
\usepackage{cite}
\usepackage{amsmath,amssymb,amsfonts}
\usepackage{algorithmic}
\usepackage{graphicx}
\usepackage{textcomp}
\usepackage{xcolor}
\usepackage{comment}
\def\BibTeX{{\rm B\kern-.05em{\sc i\kern-.025em b}\kern-.08em
    T\kern-.1667em\lower.7ex\hbox{E}\kern-.125emX}}
\begin{document}

\title{Low-Cost GNSS Anti-Jamming Through 2-Bit Phase Shift Beamforming with Machine Learning \\
}

\author{\IEEEauthorblockN{Burak Soner}
\IEEEauthorblockA{\textit{sobu Labs} \\
Ankara, Türkiye \\
burak@sobulabs.com}
\and
\IEEEauthorblockN{Ekin Uzun}
\IEEEauthorblockA{\textit{EDGE Microwave} \\
Istanbul, Türkiye \\
ekin.uzun@edgemicrowave.com}
\and
\IEEEauthorblockN{Can Aksoy}
\IEEEauthorblockA{\textit{EDGE Microwave} \\
Istanbul, Türkiye \\
can.aksoy@edgemicrowave.com}
}

\maketitle
\IEEEpubidadjcol

\begin{abstract}
We investigate low-cost GNSS anti-jamming using beamforming with inexpensive 2-bit phase shifters, constraining each complex array weight to one of four QPSK phase states (real/imaginary = -1 or +1). This severe quantization sharply limits the beampattern solution space, making conventional real-valued beamforming and naive weight quantization highly suboptimal. We formulate a discrete optimization that trades interference suppression against satellite-direction gain, and benchmark known combinatorial optimization methods across array sizes and interference conditions. Simulations show that performance improves with array size, with oracle and greedy search achieving up to 34 dB nulling, but oracle incurs exponential latency and greedy sampling is stochastic. To obtain deterministic low-latency performance, we propose an ML-aided method based on gradient-boosted decision trees followed by local search, which performs similar to the oracle for larger arrays at fixed latency. We further validate the approach experimentally using a fully digital emulation of the QPSK oracle beamformer and compare against a GNSS receiver without beamforming capability. Under mild jamming (J/S $\approx$ 44 dB) both receivers maintain adequate tracking, with QPSK yielding a 4.2 dB higher average C/N$_0$ (37.3 vs. 33.1 dB-Hz). Under moderate and strong jamming (J/S $\approx$ 62-70 dB) the benefit is substantial. At J/S = 70 dB the unprotected receiver degrades to near tracking limits (avg C/N$_0$ = 9.3 dB-Hz) while the QPSK oracle sustains an average C/N$_0$ of 20.8 dB-Hz. These results confirm that 2-bit phase-shift beamforming provides considerable anti-jamming benefit over a standard GNSS receiver, motivating further research on oracle-level practical methods.
\end{abstract}


\section{Introduction}
GNSS receivers increasingly face risk from high-power jammers that can cause loss of navigation. Controlled radiation pattern antenna (CRPA) arrays mitigate such threats by adaptive beamforming, placing deep nulls on the array pattern toward interferers while preserving gain toward satellites \cite{b1}. CRPAs rely on calibrated multi-channel RF and heavy digital processing, driving cost, size, and power, making the search for efficient algorithms worthwhile \cite{b2}.

This work targets a practical point in this space: extremely low-cost anti-jamming with beamforming using only 2-bit phase-only array weights, realizable with inexpensive analog phase shift based CRPAs. Complex array weights are constrained to a QPSK phase dictionary (real and imaginary values are either +1 or -1), enabling implementations using commodity phase shifters or switch-and-delay networks. This approach has been studied for other applications such as telecommunications \cite{b3,b4}, however its quantitative analysis for GNSS interference mitigation was not investigated.

Recent work on low-resolution beamforming covers a general analysis of gain under nonideal phase constraints \cite{b10}. On the GNSS anti-jamming (AJ) side, \cite{b11} studies jammer suppression with a 4-element discrete beamformer using 6-bit phase shifters. Their method first does an exhaustive search over the top 2 bits, followed by neighborhood refinement over the remaining bits based on measured output power. Alternatively, we investigate a direct 2-bit QPSK beamformer for very low-cost use cases. We further explore the use of machine learning (ML) to find low-latency solutions that properly scale for $>$4 antennas, benchmark all methods, and provide experimental validation. A brief comparison of this paper with the closest related works is given in Table~\ref{tab:relwork}. The drawback of this QPSK approach is that heavy quantization sharply restricts the set of realizable beampatterns and turns beamforming into a discrete, nonconvex search problem. Conventional beamforming solutions are derived over a continuous weight space, becoming highly suboptimal when naively quantized to the QPSK constraint set. The constrained problem must also preserve the desired satellite response, since interference suppression alone can collapse useful GNSS signal power and harm tracking. This paper therefore formulates a discrete objective that explicitly trades interference power suppression against gain toward a desired satellite direction, and treats the resulting QPSK-constrained weight selection as a structured combinatorial optimization problem. 

\begin{table}[t]
\caption{Brief comparison with related work.}
\label{tab:relwork}
\centering
\begin{tabular}{lllccc}
\hline
Ref. & Bf. & Quantization & GNSS AJ & Benchmk. & Exp. Valid. \\
\hline
\cite{b1} & \checkmark &  &  &  &  \\
\cite{b3,b4} & \checkmark & \checkmark &  &  &  \\
\cite{b11} & \checkmark & \checkmark ~(6-bit) & \checkmark &  & \checkmark \\
Ours & \checkmark  & \checkmark ~(2-bit, ML) & \checkmark & \checkmark & \checkmark \\
\hline
\end{tabular}
\vspace{-10px}
\end{table}

Our contributions are as follows:

\begin{itemize}
    \item We benchmark known methods under various array sizes and simulated interference conditions: exhaustive oracle selection and greedy sampling. These baselines quantify the achievable null depth and the satellite-signal loss induced by the QPSK constraint.
    \vspace{5px}
    \item We investigate an ML-aided algorithm based on gradient-boosted decision trees (GBDT), targeting oracle-level accuracy and lower latency with deterministic output.
    \vspace{5px}
    \item We experimentally test the QPSK beamforming constraint via fully digital emulation and compare its performance to a GNSS receiver without beamforming capability.
\end{itemize}

The results identify the operating regimes where the proposed QPSK beamforming can be practically useful and quantify the benefit it provides over a receiver without spatial filtering. The paper is organized as follows: Section~II defines the system model and the discrete optimization objective. Section~III describes the proposed low-cost QPSK beamforming and the benchmarked methods, including the ML-aided method. Section~IV presents simulation results across array sizes and interference scenarios. Section~V reports experimental results using a real-time digital emulation and compares performance against a GNSS receiver without beamforming. Section~VI concludes and summarizes the main takeaways.

\section{System Model and Problem Definition}

We consider an $N$-element CRPA receiver operating over a complex baseband around the GNSS carrier. Let $x[k]\in\mathbb{C}^{N}$ denote the array sample at time index $k$, after downconversion. A standard sample model is
\begin{equation}
x[k] = a_g\, s[k] + a_j\, u_j[k] + n[k],
\label{eq:receiver_model}
\end{equation}
where $s[k]$ is a desired GNSS component, $u_j[k]$ is the jammer component, $a_g\in\mathbb{C}^{N}$ is the desired steering vector, $a_j\in\mathbb{C}^{N}$ is the interference steering vector, and $n[k]$ is noise. We denote the sample covariance matrix computed over a window of $K$ samples as
\begin{equation}
\hat R = \frac{1}{K}\sum_{k=1}^{K} x[k]x[k]^H,
\label{eq:scm}
\end{equation}
where $^H$ denotes Hermitian transpose. A linear beamformer combines the array channels using $w\in\mathbb{C}^{N}$,
\begin{equation}
y[k] = w^H x[k].
\label{eq:bf_output}
\end{equation}

The real-valued optimal beamformer (Capon beamformer) computes continuous (unquantized) weights that minimize output power while enforcing a distortionless response toward the desired steering vector $a_g$:
\begin{equation}
\min_{w\in\mathbb{C}^{N}} \;\; w^H \hat R\, w
\quad \text{s.t.}\quad
w^H a_g = 1.
\label{eq:capon_opt}
\end{equation}
Assuming $\hat R$ is nonsingular (or after diagonal loading) \cite{b5}, the closed-form solution is: 
\begin{equation}
w_{\text{Capon}} = \frac{\hat R^{-1} a_g}{a_g^H \hat R^{-1} a_g}.
\label{eq:capon_sol}
\end{equation}
This formulation preserves unity gain in the desired direction while suppressing energy arriving from directions emphasized by $\hat R$. However, the final weights are typically adjusted or optimized to a fixed $L_p$-norm in practice, attenuating the satellite power up to 3-4 dB for the sake of containing the overall dynamic range.

\subsection{Performance Metrics}
The main objective is to suppress interferer power while keeping satellite-direction response distortionless. Interference suppression is measured by the beamformer output power,
\begin{equation}
P_{\text{out}}(w) \triangleq \mathbb{E}\{|y|^2\} \approx w^H \hat R\, w.
\label{eq:pout}
\end{equation}
Desired-signal preservation is quantified by the beampattern gain toward the satellite direction,
\begin{equation}
G_g(w) \triangleq |w^H a_g|^2.
\label{eq:ggain}
\end{equation}
In our benchmarks, we report null depth (reduction of interference output power) alongside satellite-direction loss (degradation in $G_g(w)$ relative to a reference).

\subsection{The 2-Bit Phase Shift Beamforming Problem}
To enable very low-cost channel combination with 2-bit phase shifters or delay networks, we propose constraining each weight to a QPSK phase dictionary
\begin{equation}
\mathcal{Q}=\{1+j,\; 1-j,\; -1+j,\; -1-j\},
\label{eq:qpsk_dict}
\end{equation}
and enforce constant-modulus normalization by
\begin{equation}
w \in \mathcal{Q}^{N}/\sqrt{2N}.
\label{eq:qpsk_constraint}
\end{equation}
This restriction yields a discrete, nonconvex design problem: the continuous solution in \eqref{eq:capon_sol} is no longer feasible, and naive quantization of $w_{\text{Capon}}$ is generally suboptimal.

Since an analytical solution to this combinatorial optimization problem is not straightforward like in the continuous case over Lagrange multipliers, we define the QPSK-constrained beamforming problem as a multi-objective optimization problem, selecting $w$ to trade jammer suppression against satellite signal preservation. In weighted-sum form,
\begin{equation}
\max_{w \in \mathcal{Q}^{N}/\sqrt{2N}}
\;\; \alpha\, |w^H a_g|^2 \;-\; (1-\alpha)\, w^H \hat R\, w,
\label{eq:qpsk_tradeoff}
\end{equation}
where $\alpha\in[0,1]$ controls the trade-off. The remainder of the paper evaluates practical methods for optimizing \eqref{eq:qpsk_tradeoff} and quantifies the resulting performance vs. latency trade-off.

\section{Proposed Low-Cost QPSK Beamforming}

This section describes the benchmarked optimization methods for the QPSK-constrained problem in \eqref{eq:qpsk_tradeoff} and the novel ML-aided approach. All methods operate on $\hat R$ and the desired steering vector $a_g$, and output a weight vector $w\in\mathcal{Q}^N/\sqrt{2N}$.

\subsection{Baseline: Naive Quantization}
A natural baseline is to compute the continuous solution \eqref{eq:capon_sol} and then quantize each entry to the nearest QPSK phase state (i.e., real and imaginary values get quantized to the nearest of -1 and +1). This approach is computationally light and serves as a lower-bound baseline, but it ignores the discrete nature of \eqref{eq:qpsk_tradeoff} and is often far from optimal, especially when fine phase resolution is required.

\subsection{Oracle Upper Bound}
For reference, we define an exhaustive oracle that evaluates the objective in \eqref{eq:qpsk_tradeoff} over all $4^{N}$ candidates and returns the optimum, with respect to an empirical selection of $\alpha=0.01$. Although its latency grows exponentially with $N$, the oracle is practical for a limited number of antennas, e.g., in \cite{b11} for a 4-element array, whose 6-bit method collapses to an oracle search if a QPSK constraint is imposed.

\subsection{Search-Based Methods in the QPSK Space}
To bridge the gap between the naively-quantized solution and the oracle, we consider practical discrete optimization methods that directly operate on $\mathcal{Q}^N$.

\subsubsection{Greedy Random Sampling}
We sample $S$ candidates $w^{(i)}\in\mathcal{Q}^N/\sqrt{2N}$ uniformly at random and select the best according to \eqref{eq:qpsk_tradeoff}. This provides a tunable accuracy vs. latency trade-off controlled by $S$, but is fully stochastic.

\subsubsection{Local Descent by Coordinate Updates}
Starting from an initialization $s^{(0)}\in\mathcal{Q}^N$ (e.g., that of the naively-quantized solution), we perform coordinate-wise descent directly in the QPSK space. At each sweep and for each antenna index $i$, we test all QPSK symbols and select the one that minimizes the combined objective:
\begin{equation}
s_i^{(t+1)} \in \arg\min_{q\in\mathcal{Q}}
\; J_{\text{c}}\!\left(s_1^{(t+1)},\dots,s_{i-1}^{(t+1)}, q, s_{i+1}^{(t)},\dots,s_{N}^{(t)}\right),
\label{eq:coord_update}
\end{equation}
where $J_{\text{c}}(\cdot)$ is the trade-off cost used in \eqref{eq:qpsk_tradeoff}. Using the newest values for indices $<i$ and the previous values for indices $>i$ corresponds to a Gauss-Seidel sweep \cite{b8}. The algorithm repeats sweeps until no coordinate update reduces $J_{\text{c}}(\cdot)$, yielding a discrete local optimum. Each sweep requires at most $4N$ objective evaluations, and performance heavily depends on good initialization.

\subsection{Proposed ML-Aided Method: GBDT + Local Refinement}
We propose a two-stage method that mimics the oracle solution of \eqref{eq:qpsk_tradeoff} with low latency. First, a learned policy predicts a QPSK weight vector directly from $\hat R$ and steering information. Second, a short coordinate-descent refinement improves the predicted solution while keeping runtime bounded.

\subsubsection{Training Data Generation (Oracle Labels)}
Training samples are generated with inputs from $\hat R$ and steering vector samples of the received signal model in \eqref{eq:receiver_model}, and outputs from the oracle selection method, which finds the optimum solution over exhaustive search. Satellite and jammer directions as well as power ratios are randomized over feasible intervals of the assumed RF frontend. While this kind of machine learning and deep learning based beamforming techniques is not new \cite{b9}, they are typically applied in the continuous solution space regime with supervised training on a direct performance objective. Our choice of oracle-based training resembles a knowledge-distillation framework on tabular data due to the severely constrained QPSK dictionary, which motivates the use of a decision tree rather than a neural network since the problem is closer to a 4-class classification problem per antenna and the resulting inference is deterministic and lightweight.

\subsubsection{Feature Extraction}
From each pair $(\hat R,a_g)$ we form a real-valued feature vector $\phi(\hat R,a_g)$ consisting of:
(i) the lower-triangular entries of $\hat R$ (real and imaginary parts),
(ii) the diagonal entries of $\hat R$,
(iii) the eigenvalues of $\hat R$ (sorted), and
(iv) the real/imaginary parts of $a_g$ as well as its magnitude and phase per element.
This representation is fixed-length for a given array size and is directly computed from the readily-available $\hat R$.

\subsubsection{GBDT Ensemble for QPSK Weight Prediction}
We train an ensemble of $N$ gradient-boosted decision tree classifiers (XGBoost) \cite{b7}, one per antenna element. Each classifier predicts one of four QPSK symbols, i.e., a class label in $\{0,1,2,3\}$ corresponding to $\{1{+}j,\,1{-}j,\,-1{+}j,\,-1{-}j\}$. The ensemble output is mapped back to a QPSK weight vector
\begin{equation}
s_{\text{GBDT}} = \big[s_1,\ldots,s_N\big]^T \in \mathcal{Q}^N,
\end{equation}
with constant-modulus normalization applied as in \eqref{eq:qpsk_constraint}. A separate model is trained for each array configuration.

\subsubsection{Local Refinement (Coordinate Descent in QPSK Space)}
The raw ML prediction $s_{\text{GBDT}}$ is refined using a small number of coordinate-descent iterations, initialized at $s^{(0)}=s_{\text{GBDT}}$. In each coordinate update, we test all four QPSK symbols for one index $i$ and accept the symbol that decreases the combined objective, using \eqref{eq:coord_update}.

\subsubsection{Runtime Considerations}
GBDT inference has fixed cost and is well-suited to embedded processors. The refinement step adds at most $4N$ objective evaluations per sweep, with a small configured iteration budget, yielding a controllable accuracy vs. latency trade-off while remaining close to the oracle operating curve in many scenarios.

\section{Simulated Benchmark}

\subsection{Simulation Setup}
We evaluate the proposed QPSK-constrained beamforming methods using a custom-developed simulation pipeline. For a given antenna configuration, a desired GNSS baseband sequence $s[k]$ generated by \texttt{gps-sdr-sim} \cite{bg} is steered to the array, producing per-element complex baseband samples. Spatially uncorrelated receiver noise is added to adjust SNR, and a spatially coherent jammer source (full-band AWGN) is added by generating an interference waveform and steering it according to randomized interference angles. We simulate against various J/S levels as well as satellite-interferer azimuth-elevation locations in Monte-Carlo style.

We compare the following methods:
(i) \emph{Naively-Quantized}, where the continuous solution is quantized to the nearest QPSK vector;
(ii) \emph{Oracle}, exhaustive search over all $4^N$ QPSK vectors for the combined objective, as a projection of the existing work \cite{b11} onto the QPSK space;
(iii) \emph{Greedy}, random sampling of a fixed number of QPSK vectors and selecting the best objective value;
(iv) \emph{GBDT + Local Refinement}, where a per-antenna GBDT ensemble predicts an initial QPSK vector from covariance and steering features, followed by a bounded number of coordinate-descent iterations.
We report two scalar metrics per trial: the normalized beampattern gain toward the satellite direction and toward the interference direction.

\begin{figure*}[!t]
    \centering
    \includegraphics[width=0.93\textwidth]{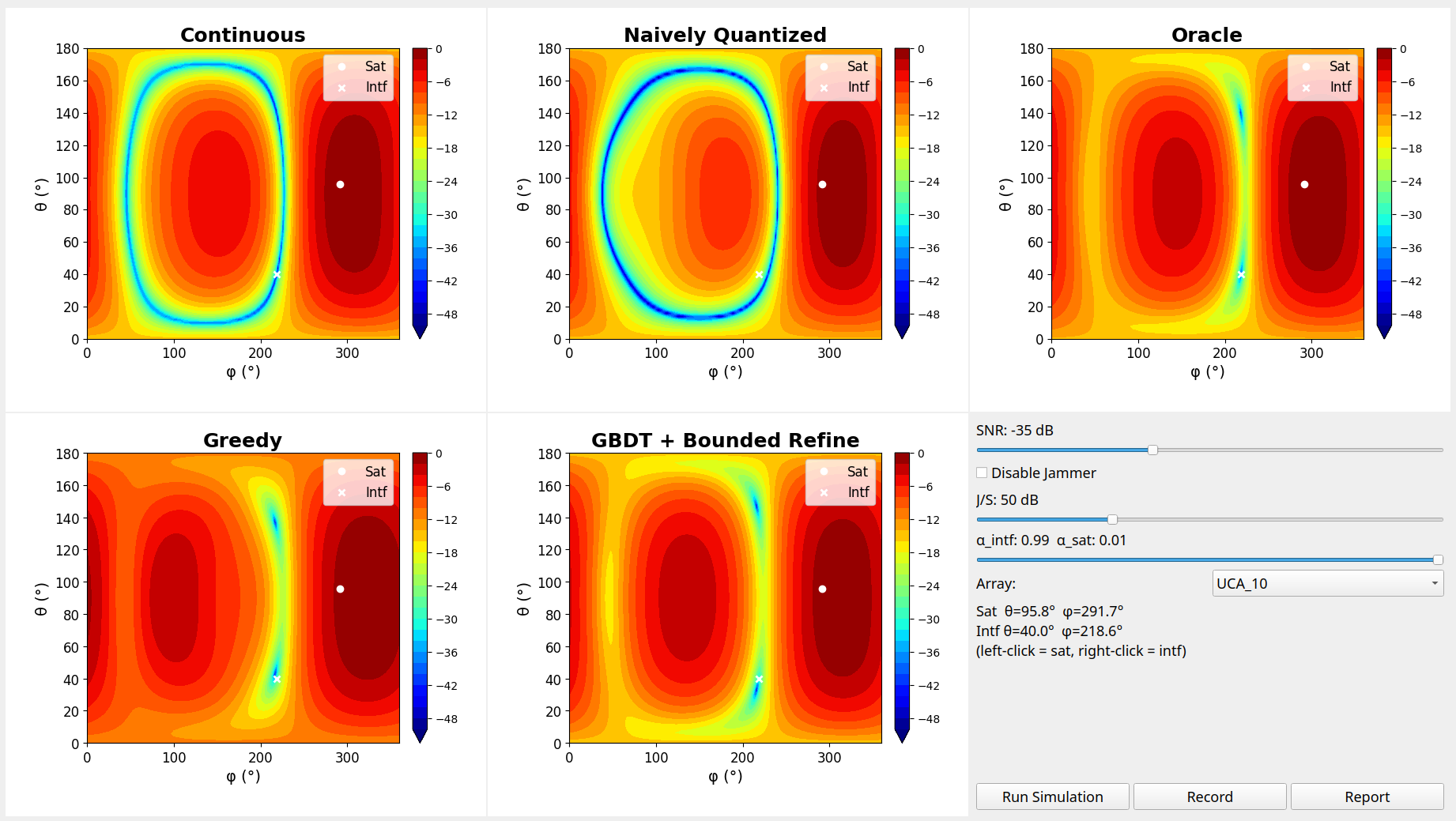}
    \caption{Representative beampattern comparison for a single randomized scenario.}
    \label{fig:beamplots}
    \vspace{10px}
\end{figure*}

\subsection{Results}
Tables~\ref{tab:sim_results_uca4}-\ref{tab:sim_results_uca10} report the satellite-direction gain, jammer-direction gain, and measured inference latency averaged over a random batch of satellite/jammer configurations and J/S levels. Larger negative jammer-direction gain indicates deeper nulling. As expected, performance improves with array size because the number of realizable QPSK beampatterns grows as $4^N$, filling more parts of the quantized solution space. The oracle, representing existing work \cite{b11} for QPSK, provides an upper bound under the QPSK constraint, while the naively-quantized solution is consistently the weakest baseline.

\begin{table}[t]
\centering
\caption{Simulated results for $N=4$.}
\label{tab:sim_results_uca4}
\begin{tabular}{lccc}
\hline
\textbf{Method} & \textbf{Sat (dB)} & \textbf{Intf (dB)} & \textbf{Infer. (ms)} \\
\hline
\textcolor{red}{Capon (continuous)}   & \textcolor{red}{-1.92} & \textcolor{red}{-156.54} & \textcolor{red}{0.12} \\
QPSK-Quantized                 & -4.07 & -20.66  & 0.16 \\
Oracle (\hspace{1sp}\cite{b11} for QPSK)   & -9.45 & -30.81  & 3.14 \\
Greedy                         & -9.45 & -30.81  & 25.31 \\
GBDT+Refine                    & -10.65 & -31.93 & 7.74 \\
\hline
\end{tabular}
\end{table}

\begin{table}[t]
\centering
\caption{Simulated results for $N=8$.}
\label{tab:sim_results_uca8}
\begin{tabular}{lccc}
\hline
\textbf{Method} & \textbf{Sat (dB)} & \textbf{Intf (dB)} & \textbf{Infer. (ms)} \\
\hline
\textcolor{red}{Capon (continuous)}   & \textcolor{red}{-1.83} & \textcolor{red}{-156.38} & \textcolor{red}{0.13} \\
QPSK-Quantized                 & -2.63 & -13.47  & 0.18 \\
Oracle (\hspace{1sp}\cite{b11} for QPSK)   & -2.26 & -34.50  & 486.35 \\
Greedy                         & -4.50 & -32.14  & 24.03 \\
GBDT+Refine                    & -2.70 & -26.79  & 15.81 \\
\hline
\end{tabular}
\end{table}

\begin{table}[t]
\centering
\caption{Simulated results for $N=10$.}
\label{tab:sim_results_uca10}
\begin{tabular}{lccc}
\hline
\textbf{Method} & \textbf{Sat (dB)} & \textbf{Intf (dB)} & \textbf{Infer. (ms)} \\
\hline
\textcolor{red}{Capon (continuous)}   & \textcolor{red}{-0.43} & \textcolor{red}{-77.96} & \textcolor{red}{0.12} \\
QPSK-Quantized                 & -0.66 & -17.70 & 0.16 \\
Oracle (\hspace{1sp}\cite{b11} for QPSK)   & -0.90 & -34.91 & 7557.90 \\
Greedy                         & -0.81 & -28.90 & 29.09 \\
GBDT+Refine                    & -1.05 & -29.49 & 18.71 \\
\hline
\end{tabular}
\end{table}

Across all tested array sizes, greedy sampling can statistically approach the oracle given enough samples, but it remains stochastic (can fail miserably once in a while) and incurs latency proportional to the number of samples evaluated (currently 100 out of the 256, 65536 and 1048576 samples within the $N=4,8,10$ solution spaces, respectively). The proposed GBDT + refinement method achieves better accuracy vs. latency trade-off at larger $N$, providing near-oracle behavior in many trials with fixed inference latency, and deterministic output.

In addition to beampattern metrics, the Tables also quantify inference latency on a CPU representative of CRPA platforms, so the relative compute costs are indicative of practical deployment. The continuous beamformer simulation has very low latency because it requires a single linear solve (or matrix inversion) per covariance update. In contrast, QPSK-constrained methods have to evaluate the same matrix-vector quadratic forms repeatedly: oracle incurs exponential cost due to $4^N$ candidate evaluations, while greedy and coordinate-descent scale with the number of sampled candidates or local updates. The ML-aided method reduces this cost by replacing most arithmetic-heavy search with a fixed-cost GBDT inference stage, which is dominated by threshold comparisons along tree paths, followed by a short, bounded local refinement. This comparison-based structure is attractive for efficient embedded realization and motivates hardware-oriented implementations of the learned policy.

Fig.~\ref{fig:beamplots} shows a comparison of the achieved elevation-azimuth beampatterns for a single scenario for $N=10$, including regular real-valued beamformer simulation without the QPSK constraint. The oracle achieves the deepest jammer null while preserving the desired-direction response. The proposed GBDT + refinement closely matches the oracle beampattern, including the location and depth of the null, while avoiding the stochastic and iterative nature of greedy search.

\begin{figure*}[t]
    \centering
    \includegraphics[width=0.99\textwidth]{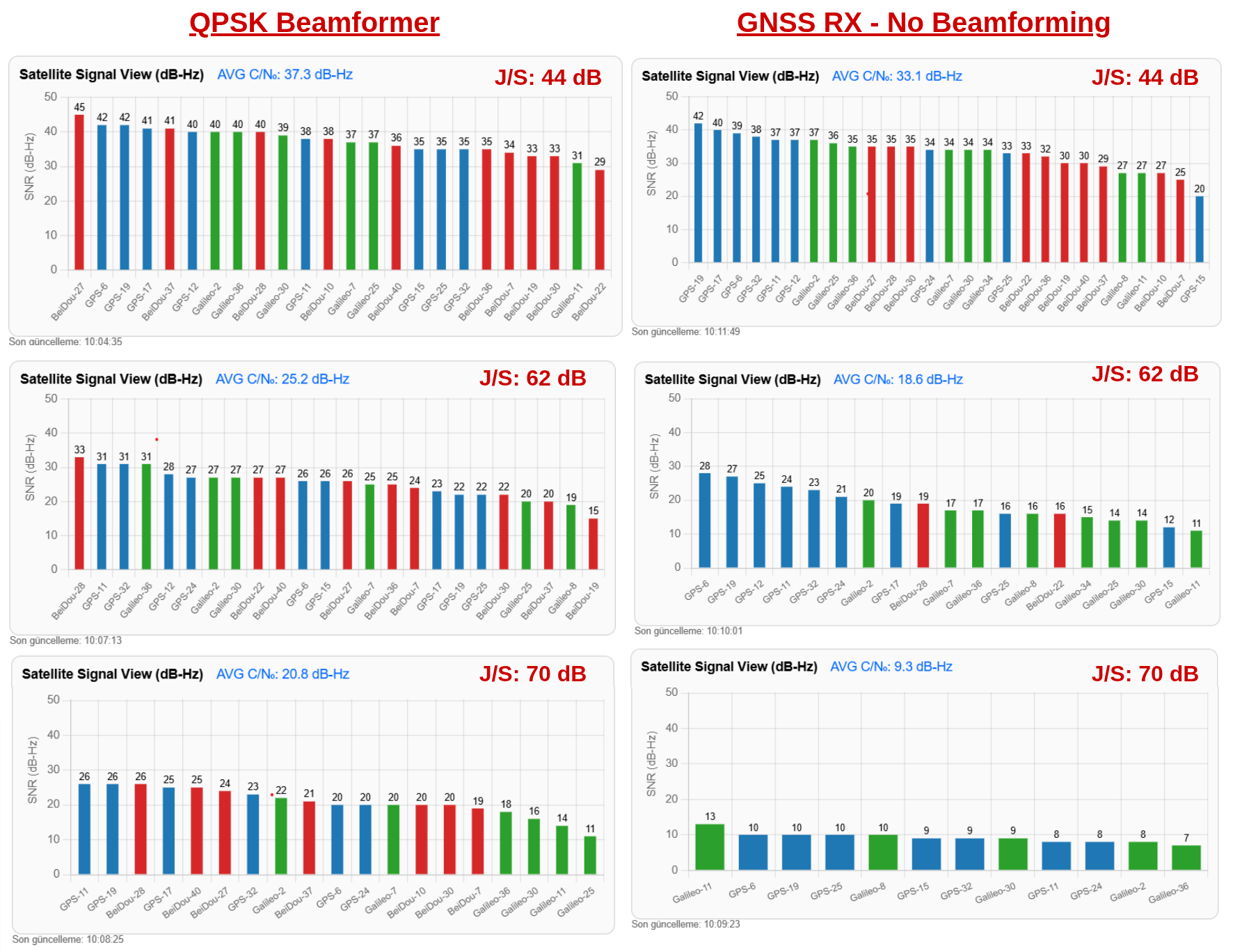}
    \caption{Experimental GNSS C/N$_0$ results comparing the proposed QPSK oracle beamformer against a GNSS receiver without beamforming under mild, moderate and high jamming levels (J/S $\approx$ 44, 62, and 70 dB). Average C/N$_0$ under no jamming for this location / orientation is $\approx$~37.6 dB-Hz, indicating a challenging reception condition due to setup isolation.}
    \label{fig:experimental}
\end{figure*}

\section{Experimental Results}
This section evaluates the practical impact of QPSK-constrained (2-bit phase shift) beamforming on a real GNSS receiver. We compare (i) a GNSS receiver without beamforming capability, representing the baseline without spatial filtering, and (ii) the proposed QPSK-constrained approach using the \emph{oracle} solution of \eqref{eq:qpsk_tradeoff}, which represents the best achievable performance under the QPSK alphabet. We use the oracle (rather than learned / searched methods) to demonstrate the fundamental performance gain achievable under the 2-bit QPSK constraint, motivating future research on efficient methods that can find oracle-level QPSK combinations with low latency.

\subsection{Setup and Procedure}
Experiments are conducted in an isolated and controlled environment with the receiver held static (fixed position and orientation) and a single full-band fast-sweeping non-linear chirp jammer. For each jamming condition, we record tracking performance with either the QPSK oracle beamformer or without any beamforming. The QPSK mode is realized via a fully digital emulation of the constrained beamformer at $N=8$, isolating the algorithmic impact of 2-bit QPSK beamforming from RF non-idealities that would arise in an external phase-shifter network, demonstrating an upper bound on the achievable performance. We evaluate receiver robustness under interference using GNSS tracking observables, primarily reported C/N$_0$ per tracked satellite and the number of tracked satellites.

\subsection{Results and Discussion}
Fig.~\ref{fig:experimental} reports per-satellite C/N$_0$ for three jamming levels (J/S $\approx$ 44, 62, and 70~dB), comparing the QPSK oracle beamformer against a GNSS receiver without beamforming under the same static receiver pose and jammer configuration. The baseline no-jammer average C/N$_0$ for this configuration is $\approx$~37.6~dB-Hz with 25 satellites in view, indicating a challenging setup due to isolation.

Under mild jamming (J/S $\approx 44$~dB), both receivers continue to track most visible satellites. The QPSK oracle achieves an average C/N$_0$ of 37.3~dB-Hz compared to 33.1~dB-Hz for the receiver without beamforming, a difference of 4.2~dB. Both operate comfortably above typical tracking thresholds, suggesting that anti-jamming beamforming provides a consistent but modest benefit even at lower interference levels.

Under moderate jamming (J/S $\approx 62$~dB), the benefit of QPSK beamforming increases. The oracle achieves an average C/N$_0$ of 25.2~dB-Hz, while the no-beamforming receiver degrades to 18.6~dB-Hz, a gap of 6.6~dB. Several satellites in the no-beamforming case drop below 15~dB-Hz, increasing the risk of tracking loss, whereas the QPSK oracle maintains a majority of satellites above 20~dB-Hz.

Under strong jamming (J/S $\approx 70$~dB), the difference becomes operationally critical. The QPSK oracle sustains tracking of most satellites at an average C/N$_0$ of 20.8~dB-Hz, whereas the receiver without beamforming degrades to an average of 9.3~dB-Hz with most signals at or near the tracking threshold. This confirms that without spatial filtering the jammer overwhelms the receiver, while the QPSK oracle effectively steers a null toward the interference and sustains robust tracking.

Overall, the measurements demonstrate that 2-bit phase shift beamforming provides meaningful anti-jamming benefit relative to a standard GNSS receiver across the entire tested interference range, with the advantage growing from 4.2~dB at mild jamming to 11.5~dB in average C/N$_0$ at strong jamming (J/S = 70~dB).

\section{Conclusion}
We investigated ultra-low-cost GNSS anti-jamming using beamforming with 2-bit (QPSK) array weights to analyze the feasibility of analog CRPA architectures built from inexpensive phase shifters or switch-and-delay networks. The QPSK constraint sharply limits the realizable beampattern set, making naive quantization of continuous beamformer weights inadequate and motivating a discrete optimization that explicitly trades interference suppression against satellite-direction gain.

We benchmarked several methods for the resulting combinatorial problem, including the naively quantized solution, exhaustive oracle selection, greedy random sampling, and a simple ML-aided method based on gradient-boosted decision trees followed by bounded local refinement. In simulation, exhaustive oracle search achieved up to 34~dB jammer suppression under the QPSK constraint, while the ML-aided approach provided a favorable accuracy versus latency tradeoff with deterministic output. We also performed an experimental comparison of the QPSK oracle beamformer against a GNSS receiver without beamforming capability under identical jamming conditions. The measured average C/N$_0$ advantage was 4.2~dB at J/S $\approx 44$~dB, 6.6~dB at J/S $\approx 62$~dB, and 11.5~dB at J/S $=$ 70~dB. These results confirm that 2-bit phase shift beamforming can provide meaningful GNSS anti-jamming benefit despite the severe quantization constraint. Experimental validation was limited to a static single-jammer scenario with $N=8$. Multi-jammer cases, dynamic scenarios, and further improvements for more antennas and better accuracy - latency trade-offs remain as future work.

\section*{Declaration}
\noindent Generative artificial intelligence tools were used only for proofreading, assistance in finding references, and graphical user interface coding related to this research.

\end{document}